\documentclass[12pt]{article}

\usepackage{bbold}
\usepackage{float}
\usepackage{amsmath, amssymb, setspace,cite}
\usepackage{graphicx}

\begin{document}
\vspace*{-2.cm}
\hfill {\tt CERN-PH-TH/2012-346}

\def\thefootnote{\fnsymbol{footnote}}

\begin{center}
\Large\bf\boldmath
\vspace*{0.7cm} 
Supersymmetry confronts $B_s \to \mu^+\mu^-$:\\ Present and future status\unboldmath
\end{center}
\vspace{0.5cm}

\begin{center}
A.~Arbey$^{a,b,c,}$\footnote{Electronic address: alexandre.arbey@ens-lyon.fr},
M.~Battaglia$^{c,d,}$\footnote{Electronic address: marco.battaglia@cern.ch},
F.~Mahmoudi$^{c,e,}$\footnote{Electronic address: mahmoudi@in2p3.fr},
D.~Mart\'inez~Santos$^{f}$\footnote{Electronic address: diego.martinez.santos@cern.ch}\\[0.4cm]
\vspace{0.6cm}
{\sl $^a$ Centre de Recherche Astrophysique de Lyon, Observatoire de Lyon,\\ 
Saint-Genis Laval Cedex, F-69561, France; CNRS, UMR 5574;\\ 
Ecole Normale Sup\'erieure de Lyon, Lyon, France}\\[0.3cm]

{\sl $^b$ Universit\'e de Lyon, France; Universit\'e Lyon 1, F-69622~Villeurbanne
Cedex, France}\\[0.3cm]

{\sl $^c$ CERN, CH-1211 Geneva 23, Switzerland}\\[0.3cm]

{\sl $^d$ Santa Cruz Institute of Particle Physics, University of California,\\ 
Santa Cruz, CA 95064, USA}\\[0.3cm]

{\sl $^e$ Clermont Universit{\'e}, Universit\'e Blaise Pascal, CNRS/IN2P3,\\
LPC, BP 10448, 63000 Clermont-Ferrand, France}\\[0.3cm]

{\sl $^f$ Nikhef and VU University Amsterdam,\\
Science Park 105, NL-1098 XG Amsterdam, The Netherlands}\\

\end{center}

\renewcommand{\thefootnote}{\arabic{footnote}}
\setcounter{footnote}{0}

\vspace{0.5cm}
\begin{abstract}
The purely leptonic rare decay $B_s \to \mu^+\mu^-$ is very sensitive to supersymmetric contributions which are free from the helicity suppression of its Standard Model 
diagrams. The recent observation of the decay by the LHCb experiment and the first determination of its branching fraction motivate a review of their impact on the viable 
parameter space of supersymmetry.   
In this paper we discuss the implications of the present and expected future accuracy on BR($B_s \to \mu^+ \mu^-$) for constrained and unconstrained MSSM scenarios, in 
relation to the results from direct SUSY searches and the Higgs data at the LHC. While the constraints from BR($B_s \to \mu^+ \mu^-$) can be very important in specific SUSY 
regions, we show that the current result, and even foreseen future improvements in its accuracy, will leave a major fraction of the SUSY parameter space, compatible with 
the results of direct searches, unconstrained. We also highlight the complementarity of the $B_s \to \mu^+\mu^-$ decay with direct SUSY searches. 
\end{abstract}

\newpage

\section{Introduction}
The rare decay $B_s \to \mu^+ \mu^-$ has been recognised as a probe of new physics beyond the Standard Model (SM) and one of the high priority channels for study in the LHC $B$ physics program. Because its SM predicted rate is made very small by a helicity suppression, it may reveal the contributions of additional diagrams arising in extensions of the SM, which do not suffer from the same suppression. In particular, in supersymmetric extensions of the SM (SUSY), its decay amplitude receives an enhancement by a factor of order $\tan^3\beta$~\cite{Huang:1998vb,Hamzaoui:1998nu,Babu:1999hn}, where $\tan \beta$ is the ratio of vacuum expectation values of the two Higgs fields, and the branching fraction can be larger than in the SM by one order of magnitude, or more.
The sensitivity of BR($B_s \to \mu^+ \mu^-$) has been discussed extensively in the literature in the past years, mostly in constrained versions of the Minimal Supersymmetric Standard Model (MSSM)~\cite{Choudhury:1998ze,Huang:2000sm,Dedes:2001fv,Ellis:2005sc,Carena:2006ai,Ellis:2007fu,Heinemeyer:2008fb,Eriksson:2008cx,Alok:2009wk,Akeroyd:2010qy,Golowich:2011cx,Farina:2011bh,Akeroyd:2011kd,Mahmoudi:2012un,Fowlie:2012im,Buchmueller:2012hv,Haisch:2012re,Altmannshofer:2012ks} and, more recently, in the phenomenological MSSM (pMSSM)~\cite{Arbey:2011aa,CahillRowley:2012cb,CahillRowley:2012kx}.
The recent observation of the decay with the determination of its branching fraction by the LHCb experiment to a value very close to the SM prediction~\cite{Aaij:2012ct} excludes very large deviations, motivating a review of its implications on the viability of SUSY. In this paper, we discuss these implications in the context of constrained and unconstrained MSSM models, with R-parity and CP conservation, with an eye also on the future experimental progress of this measurement at the LHC.
We show that the constraining power of BR($B_s \to \mu^+ \mu^-$) is important in specific regions of the MSSM, but leaves substantial room for the SUSY parameters. We quantify this by studying the fraction of the MSSM model points, obtained in flat scans of the parameter space, which are compatible with the present and future BR($B_s \to \mu^+ \mu^-$) constraints in the framework of the Constrained MSSM (CMSSM) and the phenomenological MSSM (pMSSM) with 19 free parameters, and account for the results on the Higgs mass and direct SUSY searches at ATLAS and CMS. 

The paper is organised as follows. In section 2 we discuss the SM prediction for BR($B_s \to \mu^+ \mu^-$), the SUSY contributions and the current experimental results. Section 3 discusses the experimental prospects at the LHC experiments. The constraints derived are described in section 4 for the CMSSM and the more general case of the pMSSM. Conclusions are provided in section 5.

\section{Current status}

\subsection{SM prediction}

\begin{figure}[t!]
\begin{center}
\includegraphics[width=5.cm]{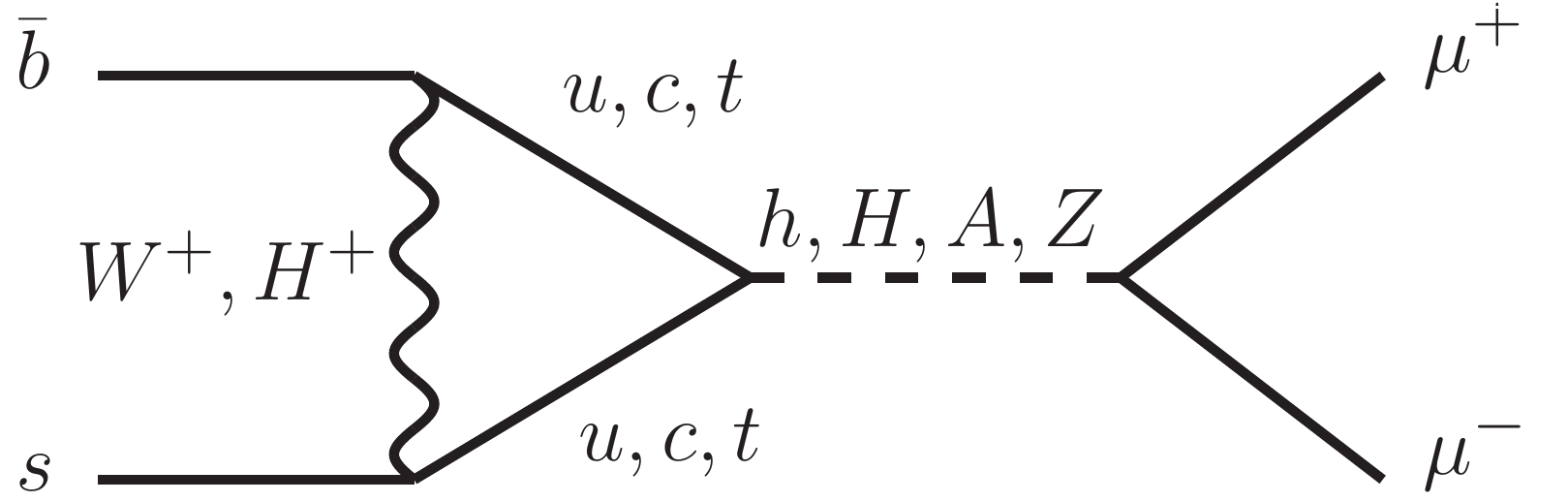}\quad\includegraphics[width=5.cm]{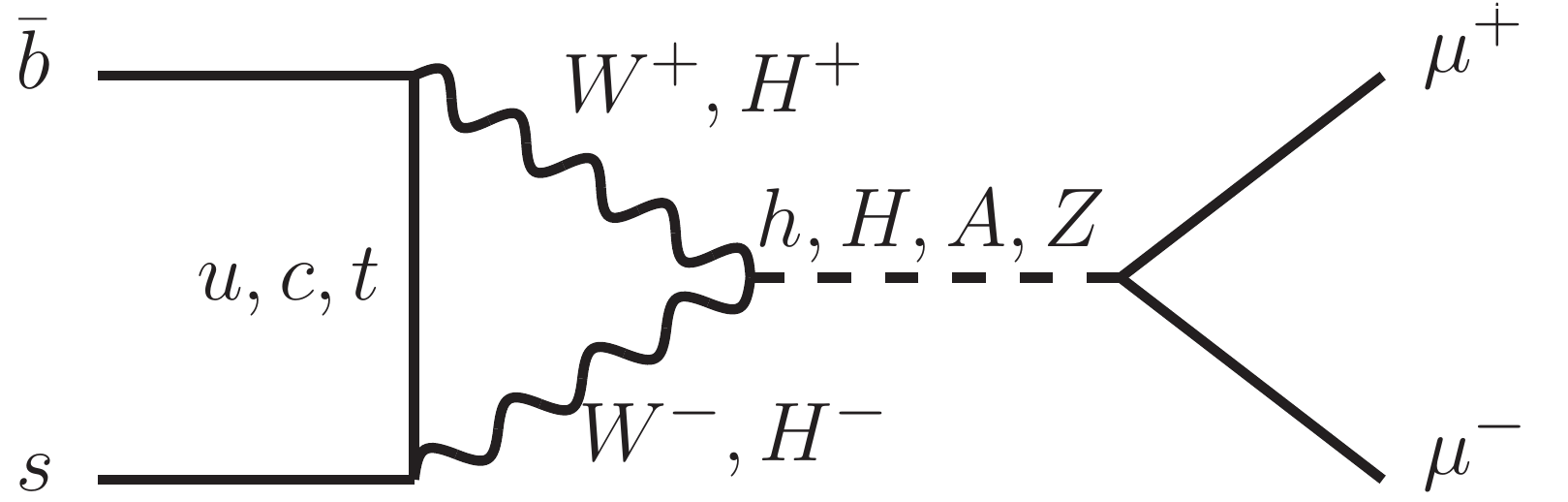}\quad\raisebox{-0.25cm}{\includegraphics[width=5.cm]{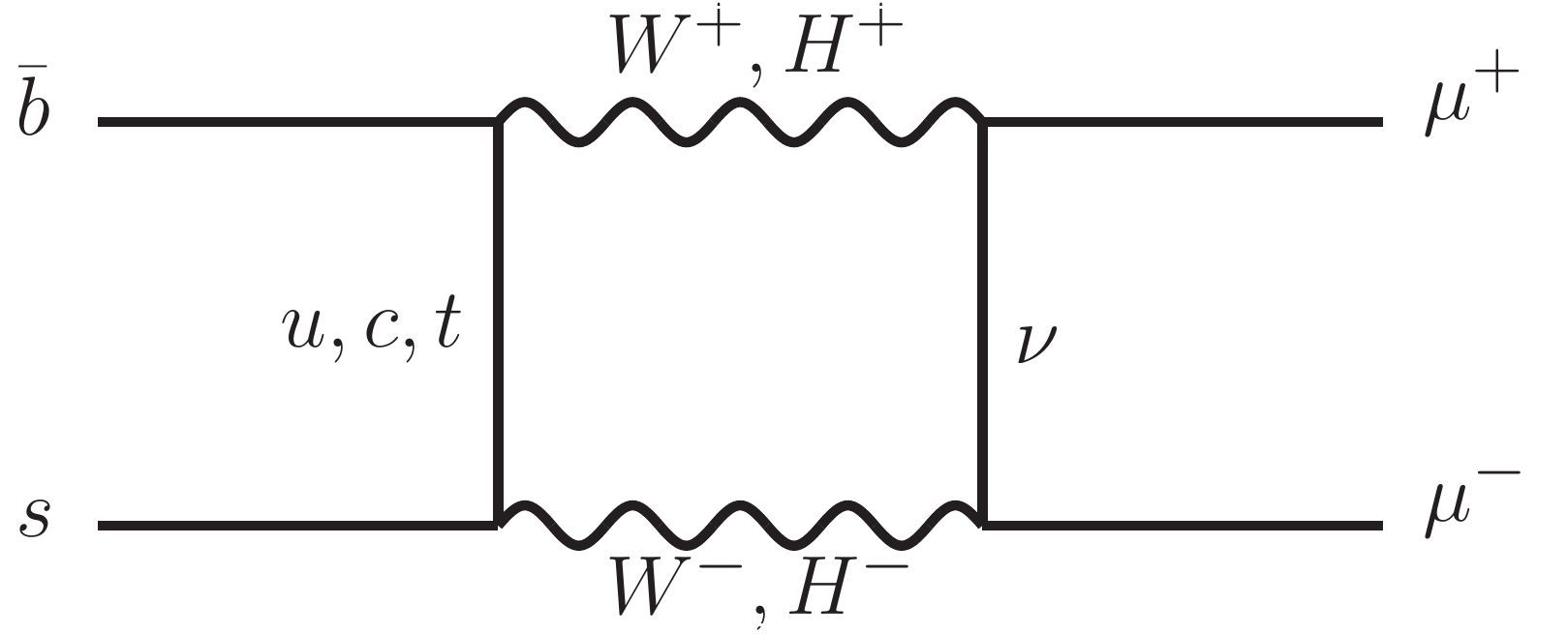}}\\[0.5cm]
\includegraphics[width=5.cm]{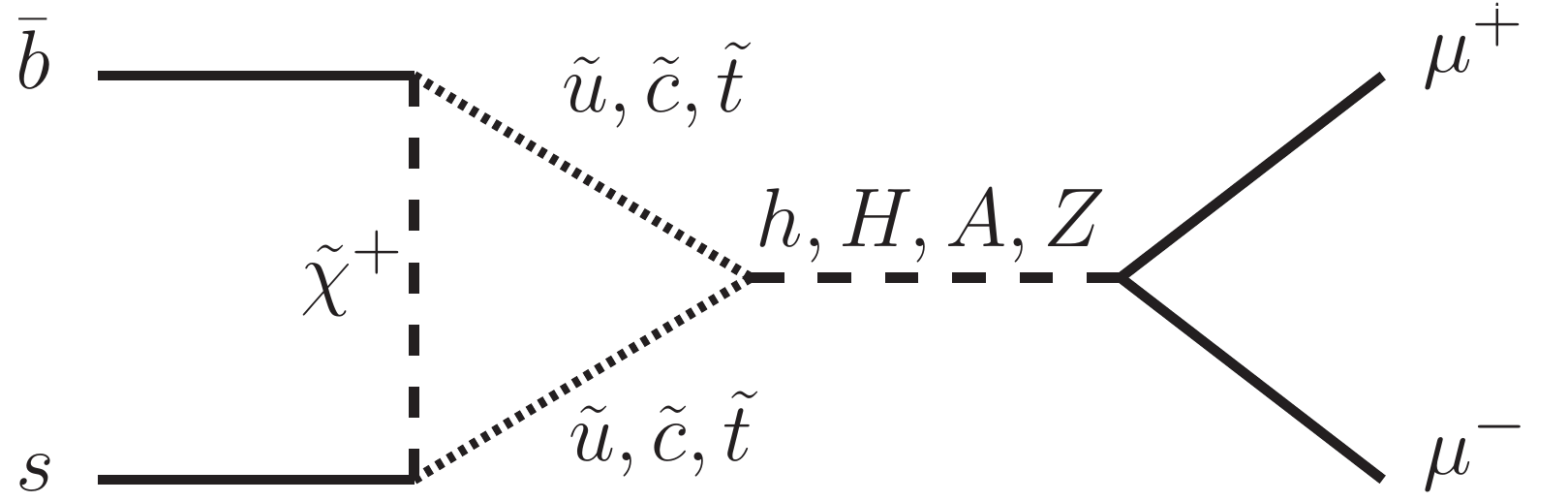}\quad\includegraphics[width=5.cm]{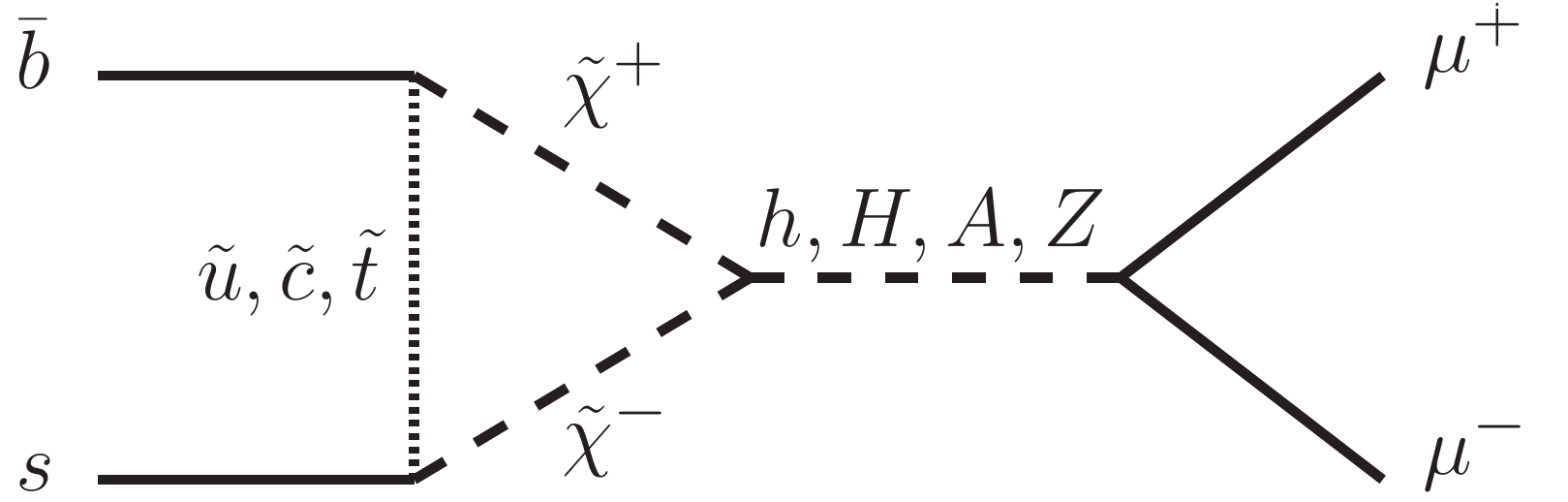}\quad\raisebox{-0.25cm}{\includegraphics[width=5.cm]{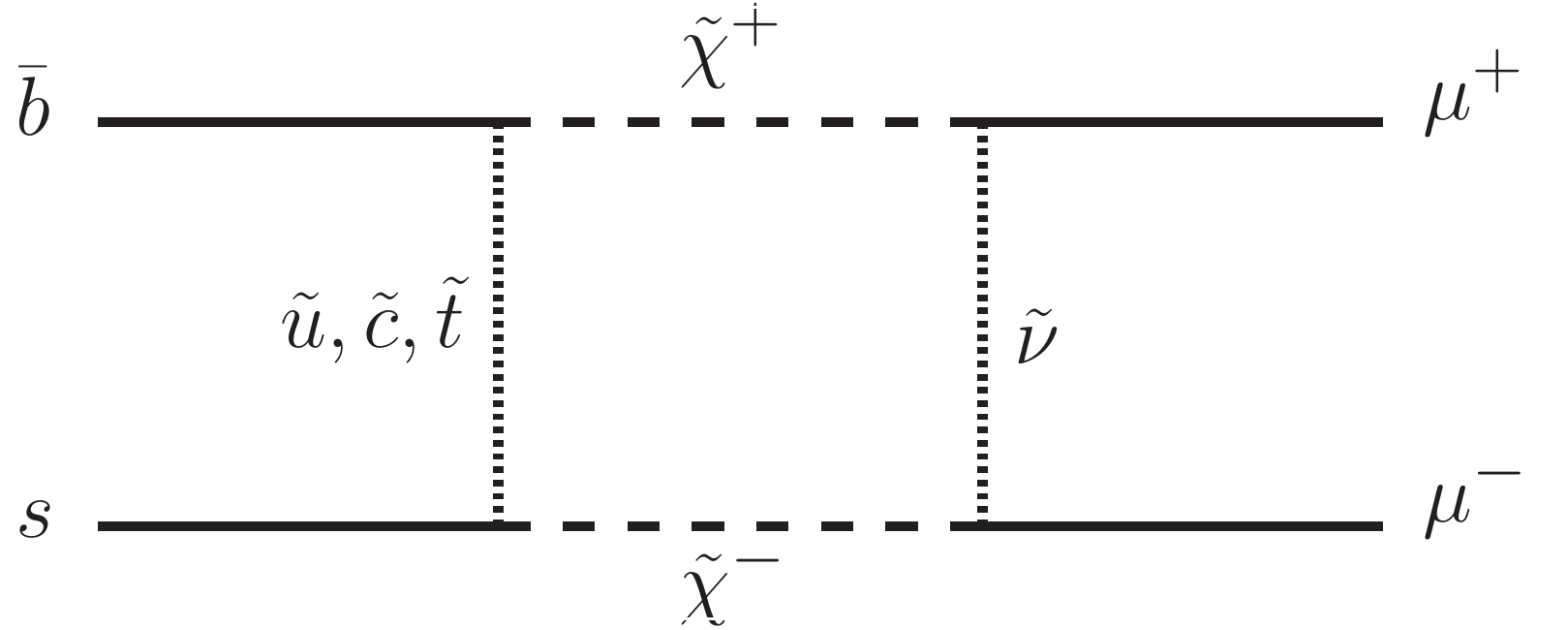}}\\[0.5cm]
\caption{Dominant $B_s \to \mu^+ \mu^-$ diagrams in the SM, 2HDM and MSSM.}
\label{fig:bsmumu}
\end{center}
\end{figure}

\begin{figure}[t!]
\begin{center}
\includegraphics[width=8.cm]{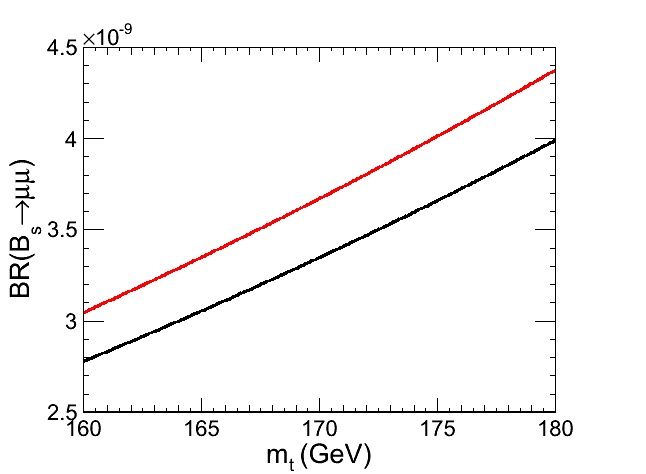}%
\caption{BR($B_s \to \mu^+ \mu^-$) vs.\ the top quark pole mass. The black (lower) line corresponds to the CP-averaged branching ratio, while the red (upper) line shows the untagged value.}
\label{fig:mtop}
\end{center}
\end{figure}

In the SM, the flavour changing neutral current (FCNC) decay $B_s \to \mu^+ \mu^-$ proceeds via $Z$ penguin and box diagrams and is helicity suppressed. The average branching fraction can be expressed as~\cite{Buchalla:1993bv,Misiak:1999yg,Bobeth:2001sq,Bobeth:2001jm,Mahmoudi:2008tp}:
\begin{eqnarray}
  \label{eq:Bsmm_formula}
\mathrm{BR}(B_s\to\mu^+\mu^-)&=&\frac{G_F^2 \alpha^2}{64\pi^3}f_{B_s}^2
m_{B_s}^3 |V_{tb}V_{ts}^*|^2\tau_{B_s}\sqrt{1-\frac{4m_\mu^2}{m_{B_s}^2}}\\
&&\times\left\{\left(1-\frac{4m_\mu^2}{m_{B_s}^2}\right)
  |C_{Q_1}-C'_{Q_1}|^2+\left|(C_{Q_2}-C'_{Q_2})+2(C_{10}-C'_{10})\frac{m_\mu}{m_{B_s}}\right|^2\right\}\,,\nonumber  
\end{eqnarray}
where $f_{B_s}$ is the $B_s$ decay constant, $m_{B_s}$ is the $B_s$ meson mass and $\tau_{B_s}$ is its mean lifetime. $C_{Q_1}$ and $C_{Q_2}$ are the Wilson coefficients of the semileptonic scalar and pseudo-scalar operators\footnote{Note that $C_{Q_{1,2}} = m_b\, C_{S,P}$.}, and $C_{10}$ 
the axial semileptonic Wilson coefficient. The $C'_{i}$ terms correspond to the chirality flipped coefficients. In the SM, only $C_{10}$ is non-vanishing and it gets its largest contributions from a $Z$ penguin top loop 
(75\%) and a $W$ box diagram (24\%) (see Fig.~\ref{fig:bsmumu}). 
The SM expected value is evaluated using $m_b^{\overline{MS}}(m_b)= (4.18 \pm 0.03)$~GeV and $m_t^{pole}= (173.5 \pm 0.6 \pm 0.8)$~GeV~\cite{Beringer:1900zz}, corresponding 
to $C_{10}=-4.16 \pm 0.04$, from which the following SM prediction for the branching fraction is derived \cite{Mahmoudi:2012un}:
\begin{equation}
\label{eq:1}
\mathrm{BR}(B_s\to \mu^+ \mu^-)
= (3.53\pm0.38) \times 10^{-9} \;,
\end{equation}
where we used the numerical values of $m_{B_s} = (5.36677 \pm 0.00024)$ GeV, $|V_{tb}V_{ts}^*|=0.0404 \pm 0.0011$, 
$\tau_{B_s}=(1.497 \pm 0.015)$~ps~\cite{Beringer:1900zz,Amhis:2012bh} and $f_{B_s} = (234 \pm 10)$ MeV. The value of $f_{B_s}$ is extracted from the average of the 
lattice results reported by the ETMC-11 \cite{Dimopoulos:2011gx}, Fermilab-MILC-11~\cite{Bazavov:2011aa,Neil:2011ku} and HPQCD-12~\cite{Na:2012kp} 
Collaborations and represents the dominant source of systematic uncertainty (8.7\%) in the SM prediction. The top mass determination and the choice of the renormalisation scheme for its running have an important impact on the evaluation of the $B_s \to \mu^+ \mu^-$ branching fraction, as discussed in~\cite{Buras:2012ru}. The effect is illustrated in Fig.~\ref{fig:mtop} where we show the SM central value for BR($B_s \to \mu^+ \mu^-$) as a function of the top pole mass value. A change of $\pm$2~GeV in the top mass corresponds to a $\pm 10^{-10}$ change in the branching fraction value.
Other sources of uncertainty include the choice of scale for the calculation of the fine-structure constant and parametric uncertainties. 
Adding all these uncertainties in quadrature, a total theoretical uncertainty of 11\% is estimated. 

\subsection{SUSY contributions}

The $B_s\to\mu^+\mu^-$ decay may receive very large enhancements within specific extensions of the SM. In particular, in the MSSM the Higgs-mediated scalar FCNCs do not suffer from the same helicity suppression as the SM diagrams, thus leading to possible drastic enhancements at large values of $\tan \beta$~\cite{Huang:1998vb,Hamzaoui:1998nu,Babu:1999hn}. 
In this case, the $C_{Q_1}$, $C_{Q_2}$ coefficients give the dominant contributions. For positive values of $C_{Q_2}$ the interference with the term proportional to $\left (C_{Q_1}^2 + C_{Q_2}^2 \right )$ is destructive. The upper bound on BR($B_s \to \mu^+ \mu^-$) is more easily evaded or, conversely, an appropriate pseudo-scalar contribution may lead to a suppression of this decay mode to rates below the SM expectation. 
In the MSSM, the largest contribution to $C_{Q_1}$ and $C_{Q_2}$, in the large $\tan\beta$ region, reads~\cite{Babu:1999hn,Haisch:2012re}:
\begin{equation}
\label{eq:CQ}
C_{Q_1} \approx -C_{Q_2} \approx -\mu A_t \, \frac{\tan^3\beta}{(1+\epsilon_b \, \tan\beta)^2} \; \frac{m_t^2}{m_{\tilde t}^2} \, \frac{m_b  m_\mu}{4\sin^2\theta_W M_W^2 M_A^2} \, f (x_{\tilde t \mu})  \;,
\end{equation}
where $x_{\tilde t \mu} = m_{\tilde t}^2/\mu^2$, with $m_{\tilde t}$ the geometric average of the two stop masses, and 
\begin{equation}
f(x) =-\frac{x}{1-x} - \frac{x}{(1-x)^2} \, \ln x \;.
\end{equation}
The $\epsilon_b$ correction parametrises loop-induced non-holomorphic terms that receive their main contributions from higgsino and gluino exchange. 
Since $f(x) >0$, the sign of $C_{Q_1}$ is opposite to that of the $\mu A_t$ term. 
Here, Eq.~(\ref{eq:CQ}) is given for purely illustrative purposes; in our numerical analysis we employ the result of a full calculation, which includes all relevant contributions. It must be pointed out that, whereas the MSSM may have a spectacular impact on the $B_s\to\mu^+\mu^-$ process, it is equally possible to effectively suppress the SUSY contributions by moving to regions of intermediate $\tan \beta$ values and/or large masses of the pseudo-scalar Higgs boson $A$. In such cases, the branching fraction does not deviate from its SM prediction, effectively preventing this decay from probing parts of the supersymmetric parameter space. 

\subsection{Experimental results}

The $B_s\to\mu^+\mu^-$ decay has been the target of a dedicated effort at the Tevatron and the LHC. To date, the most constraining upper limit obtained by a single experiment comes from LHCb~\cite{Aaij:2012ac}, $\mathrm{BR}(B_s\to\mu^+\mu^-) < 4.5 \times 10^{-9}$ at 95\% C.L., 
based on 1.0~fb$^{-1}$ of data at 7~TeV.
Searches leading to upper limits have been carried out also by CMS~\cite{Chatrchyan:2012rg} and ATLAS~\cite{Aad:2012pn}, while the CDF Collaboration reported an excess of events over the estimated background, corresponding to a value $\mathrm{BR}(B_s\to\mu^+\mu^-)  = (1.3^{+0.9}_{-0.7})\times10^{-8}$ ~\cite{Aaltonen:2011fi}. The combination of the LHCb, ATLAS and CMS results led to an upper bound of $4.2 \times 10^{-9}$~\cite{LHCb-CONF-2012-017} in Summer 2012.

More recently, the LHCb Collaboration has announced the first evidence for this decay and measured its branching fraction~\cite{Aaij:2012ct} to be:
\begin{equation}
\mathrm{BR}(B_s\to\mu^+\mu^-) = \left(3.2^{+1.4}_{-1.2}({\rm stat})^{+0.5}_{-0.3}({\rm syst})\right)\times 10^{-9}.
\label{bsmumu_evidence}
\end{equation}
This value is in excellent agreement with the SM prediction, leading to speculations on its implications on the viability of SUSY. However, it must be noted that the 
upper limit constraint derived from this result is somehow weaker compared to those from the earlier upper limits, while it is interesting to investigate the effect of 
the lower limit from (\ref{bsmumu_evidence}).

\begin{figure}[h!]
\begin{center}
\includegraphics[width=7.cm]{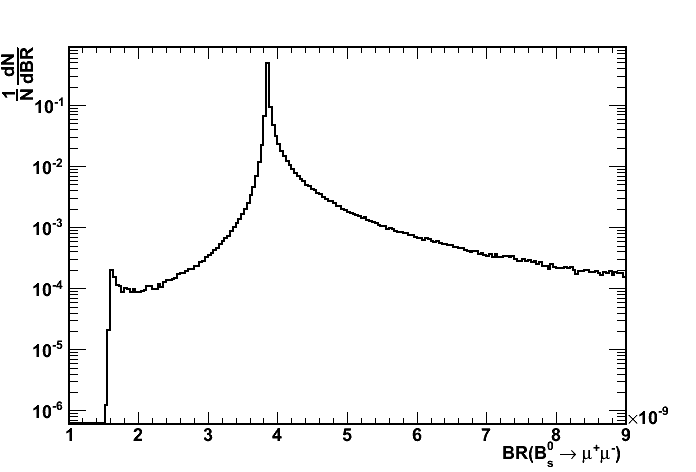}%
\caption{Distribution of BR$(B_s\to \mu^+ \mu^-)_\mathrm{untag}$ for CMSSM points. The general shape with entries at values below and above the SM expected value of 
$3.87\times 10^{-9}$ persists when restricting to the points compatible with the results of LHC SUSY searches.}
\label{fig:BR}
\end{center}
\end{figure}

Before discussing these implications, it is important to consider that the theoretical prediction of the branching fraction does not directly correspond to the quantity measured by the LHCb experiment. In fact, the theoretical predictions are CP-averaged quantities in which the effect of $B_s - \bar{B}_s$ oscillations is disregarded. On the contrary, the experimental measurement corresponds to an untagged branching fraction which is related to the CP-averaged value by the relation\cite{deBruyn:2012wj,deBruyn:2012wk}:
\begin{equation}
\mathrm{BR}(B_s \to \mu^+ \mu^-)_\mathrm{untag}=\left(\frac{1+ \mathcal{A}_{\Delta\Gamma}\,y_s}{1-y_s^2}\right) \mathrm{BR}(B_s \to \mu^+ \mu^-)\;,
\end{equation}
where
\begin{equation}
y_s \equiv \frac12 \tau_{B_s} \Delta\Gamma_s = 0.088 \pm 0.014\;,
\end{equation}
and
\begin{equation}
\mathcal{A}_{\Delta\Gamma} = \frac{|P|^2 \cos(2\varphi_P) - |S|^2 \cos(2\varphi_S)}{|P|^2 + |S|^2}\;.
\end{equation}
$S$ and $P$ are related to the Wilson coefficients by:
\begin{equation}
S = \sqrt{1-4 \frac{m_\mu^2}{M_{B_s}^2}} \, \frac{M_{B_s}^2}{2m_\mu} \, \frac{1}{m_b + m_s} \frac{C_{Q_1}-C'_{Q_1}}{C_{10}^{SM}}\;,
\end{equation}
\begin{equation}
P = \frac{C_{10}}{C_{10}^{SM}} + 
\frac{M_{B_s}^2}{2m_\mu} \, \frac{1}{m_b + m_s} \frac{C_{Q_2}-C'_{Q_1}}{C_{10}^{SM}}\;,
\end{equation}
and
\begin{equation}
\varphi_S = \arg(S)\;,\qquad \varphi_P = \arg(P)\;.
\end{equation}
The resulting untagged branching fraction can be directly compared to the experimental measurement. The SM expectation for this corrected branching fraction is:
\begin{equation}
\label{eq:1b}
\mathrm{BR}(B_s\to \mu^+ \mu^-)_\mathrm{untag} = (3.87\pm0.46) \times 10^{-9} \;.
\end{equation}
In the MSSM, the difference between the CP-averaged and the untagged values of the branching fraction depends on the specific SUSY parameters which enter $A_{\Delta\Gamma}$ but the shift is typically within $\pm$10\%. The distribution of the branching fraction values, from our CMSSM scan discussed below, is shown in Fig.~\ref{fig:BR}.

\section{Experimental prospects}
\label{sec:exp}

\begin{figure}[t!]
\begin{center}
\includegraphics[width=8.cm]{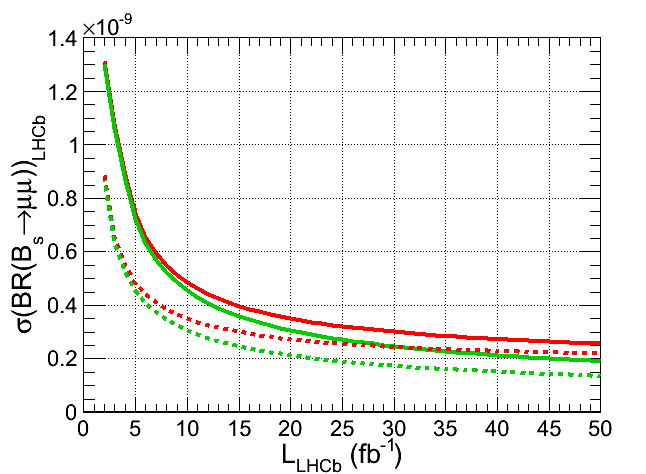}%
\caption{Expected uncertainty in the branching fraction of $B_s\to \mu^+\mu^-$ vs.\ the integrated 
luminosity recorded by LHCb (solid lines). The red (upper) line refers to an ultimate systematic uncertainty of $5\%$ and the green (lower) to an ultimate systematic uncertainty of $1\%$. 
The dashed lines show the precision of LHC combinations, assuming comparable sensitivity for the LHCb and CMS experiments in the same time period.}
\label{fig:LHCb}
\end{center}
\end{figure}

The LHC experiments, in particular LHCb, will keep improving the precision in the determination of the $B_s\to \mu^+\mu^-$ branching fraction. The latest LHCb measurement offers valid guidance for estimating the evolution of the measurement accuracy for increasing statistics. By symmetrising the statistical uncertainty of the result to $\approx 1.3$ and using Gaussian statistics,
we study the statistical accuracy as a function of the integrated luminosity. At 14~TeV centre of mass energy, the $B_s$ production cross section is approximately a factor of two larger compared to 7~TeV. The systematic uncertainties are expected to become important once the 
statistic uncertainties drop. These factors are taken into account in the following estimate:
\begin{equation}
\sigma \big(\mathrm{BR}(B_s\to \mu^+\mu^-)\big)(L) \approx \sqrt { 1.3^2\frac{2}{L} + \sigma_{syst}^2}
\end{equation}
for 7 and 8~TeV operations, and
\begin{equation}
\sigma \big(\mathrm{BR}(B_s\to \mu^+\mu^-)\big)(L) \approx \sqrt { 1.3 ^2\frac{2}{2 L -L_0} + \sigma_{syst}^2}
\end{equation}
for the 14~TeV data, where $L$ is the integrated luminosity, $L_0$ the total integrated luminosity taken at 7 and 8~TeV and $\sigma_{syst}$ the expected systematic uncertainty. With the improvements in the computing power for lattice calculations the uncertainty on $f_{B_s}$, the dominant source of theory uncertainty in the SM prediction, is likely to decrease to $\sim$1\% \cite{O'Leary:2010af}.
Fig.~\ref{fig:LHCb} shows the expected precision in BR($B_s\to \mu^+\mu^-$) as a function of the integrated luminosity for LHCb, assuming $\approx 3.5$~fb$^{-1}$ at 7 and 8~TeV, and two different scenarios for the systematic uncertainties: $5\%$ and, optimistically, $1\%$. This shows the importance of improvements in the systematic errors over a long period of time. 
The systematic uncertainty will largely depend on the accuracy available for the determination of the fragmentation function ratio $f_d/f_s$. We do not consider here improvements to the analysis and the detector performance, which are difficult to quantify at present but may lead to an additional reduction of the statistical uncertainties. The upgraded LHCb experiment plans to collect 50~fb$^{-1}$ of data after ten years of running~\cite{Bediaga:2012py},
providing an ultimate uncertainty of $\lesssim 2\times10^{-10}$.
In addition, the general purpose experiments can provide useful results and the CMS experiment has demonstrated a 
sensitivity quite close to that of LHCb. 
If this performance can be extrapolated to the future data sets, taking into account the larger event pile-up, and the higher energies, the LHC combinations 
will show improvements of $\gtrsim \sqrt{2}$ on the statistical error compared to the results of LHCb alone.
Since the systematic uncertainty on $f_d/f_s$ is common to all the experiments, it is assumed to be fully correlated in this study. 

In summary, we consider two intervals at 95\% C.L. for the branching fraction values: 
\begin{equation}
1.1 \times 10^{-9} < {\rm BR}(B_s \to \mu^+ \mu^-) < 6.4 \times 10^{-9} \label{bsmumu_current}
\end{equation}
corresponding to the current LHCb result of Eq.~(\ref{bsmumu_evidence}) and 
\begin{equation}
3.1 \times 10^{-9} < {\rm BR}(B_s \to \mu^+ \mu^-) < 4.6 \times 10^{-9} \label{bsmumu_ultimate}
\end{equation}
which represents a realistic estimate of the LHC ultimate relative accuracy of $\sim 5\%$, when including an estimated improved theory 
uncertainty of $\sim 8\%$ in the determination of the rate of this process, if the central value meets the SM prediction.

\section{Constraints in MSSM models}

We study the effect of imposing the constraints of Eqs.~(\ref{bsmumu_current}) and (\ref{bsmumu_ultimate}) on the CMSSM and pMSSM by performing 
broad scans of the model parameters and studying the fraction of points compatible with those $B_s \to \mu^+ \mu^-$ rates. The parameters are 
varied in flat scans within their ranges given below. The SUSY mass spectra are obtained with {\tt SOFTSUSY 3.3.4} \cite{Allanach:2001kg} and 
the value of BR($B_s\to \mu^+ \mu^-$) with {\tt SuperIso v3.4} \cite{Mahmoudi:2007vz,Mahmoudi:2008tp}. We select points where the lightest 
SUSY particle is the $\tilde \chi^0_1$ neutralino and which are consistent with the LEP and LEP-2 limits on SUSY particles. These points are 
referred to as ``accepted'' points in the following. Then, we test each point for compatibility with the results of the LHC SUSY and Higgs 
searches.

\subsection{CMSSM}
\label{sec:cmssm}

\begin{figure}[t!]
\begin{center}
\includegraphics[width=8.cm]{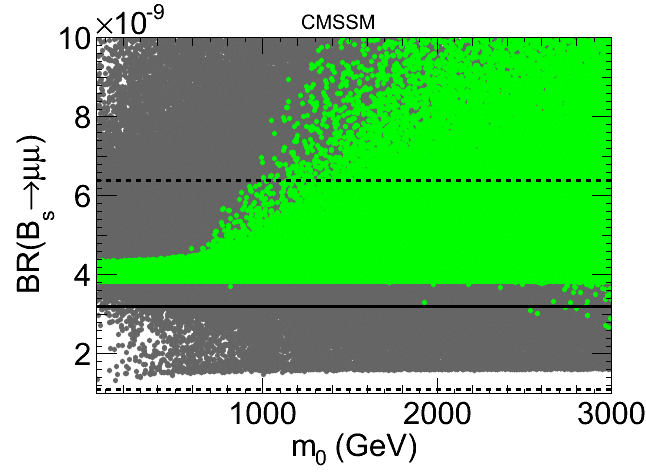}%
\includegraphics[width=8.cm]{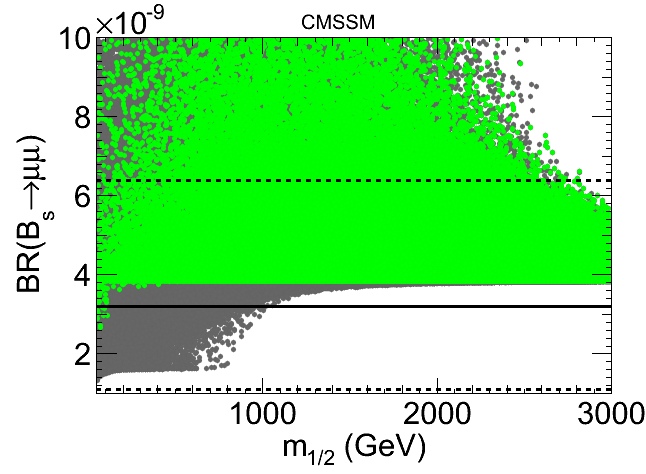}\\[0.3cm]
\includegraphics[width=8.cm]{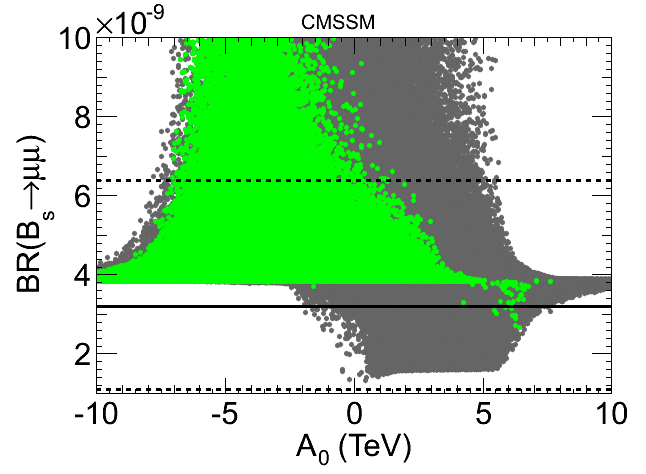}%
\includegraphics[width=8.cm]{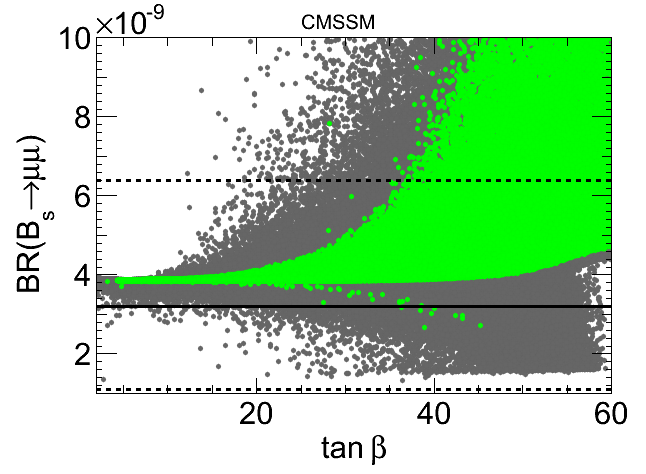}%
\caption{Untagged BR($B_s \to \mu^+ \mu^-$) vs.\ the CMSSM parameters $m_0$ (upper left), $m_{1/2}$ (upper right), $A_0$ (lower left), $\tan\beta$ (lower right). The solid line corresponds to the central value of the BR($B_s \to \mu^+ \mu^-$) measurement, and the dashed lines to the 2$\sigma$ experimental deviations. The green points are those in agreement with the Higgs mass constraint.}
\label{fig:cmssm}
\end{center}
\end{figure}

First, we consider the effect of BR($B_s \to \mu^+ \mu^-$) in the CMSSM parameter space, where we perform flat scans varying the CMSSM parameters 
in the ranges:
\begin{equation}
 m_0, m_{1/2} \in [50,3000] {\;\rm GeV};\;\tan\beta \in [1,60];\;  A_0 \in [-10,10] {\;\rm TeV};\; {\rm sign}(\mu)>0 \;.
\label{cmssm-ranges}
\end{equation}

Since the effects on BR($B_s \to \mu^+ \mu^-$) are small for negative values of the $\mu$ parameter, we choose ${\rm sign}(\mu)>0$ in the scans, which is also in better agreement with the muon $(g-2)$ constraint.

Results are given in graphical form in Fig.~\ref{fig:cmssm}, where we show the BR($B_s \to \mu^+ \mu^-$) values as functions of the four CMSSM parameters, comparing the totality of the generated points to those consistent with the lightest Higgs boson $h$ mass range, $123 < M_h < 129$~GeV~\cite{Aad:2012gk,Chatrchyan:2012gu}.
The BR($B_s \to \mu^+ \mu^-$) admits a lower value of about $1.5\times10^{-9}$, which is still larger than the present experimental lower bound derived from the LHCb measurement, so that the experimental lower limit does not yet imply the exclusion of portions of the CMSSM parameter space.
Branching fraction values below $\sim 3\times10^{-9}$ can be reached for $m_{1/2} \lesssim 1$ TeV, $0 \lesssim A_0 \lesssim 6$ TeV and $\tan\beta \gtrsim 20$. However, once the Higgs mass limits are imposed, the vast majority of the allowed points have the BR($B_s \to \mu^+ \mu^-)$ 
at values which are equal to, or larger than, the SM prediction, with the exception of a few points located in a region at large $m_0$, very small $m_{1/2}\sim 50-100$ GeV and $A_0 \sim 5$ TeV. These points are all excluded by {\it e.g.} the LEP or Tevatron direct SUSY search limits as they lead to too light gluinos and neutralinos. 

As a consequence, in the CMSSM, it is not possible to have BR($B_s \to \mu^+ \mu^-)$ smaller than the SM prediction and at the same time be in agreement with the SUSY and Higgs search results. Therefore, if in the future the central measured value of BR($B_s \to \mu^+ \mu^-$) remains close to the SM prediction, the lower bound is unlikely to have any effect on constraining the CMSSM parameter space. 

\begin{figure}[t!]
\begin{center}
\begin{tabular}{cc}
\includegraphics[width=0.45\textwidth]{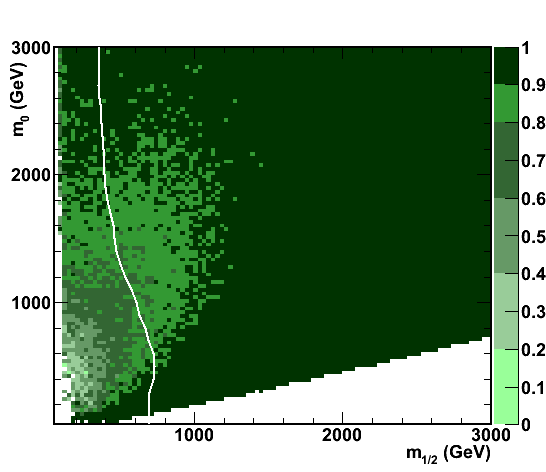} &
\includegraphics[width=0.45\textwidth]{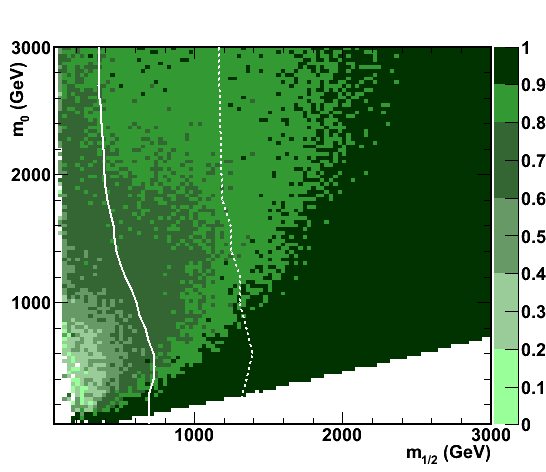} \\
\end{tabular}
\caption{Fraction of CMSSM points compatible with the current (left) and ultimate (right) 95\% C.L. constraints on 
BR($B_s \to \mu^+ \mu^-$) in the $(m_{1/2},m_0)$ parameter plane. The continuous line shows the parameter region excluded 
by the ATLAS SUSY searches at 8 TeV with 5.8~fb$^{-1}$ of data (from \cite{ATLAS-CONF-2012-109})and the dotted line the  
reach estimated by CMS for searches at 14~TeV with 300~fb$^{-1}$ (from~\cite{Abdullin:1998pm}).}
\label{fig:cmssm_exclusion}
\end{center}
\end{figure}

Figure~\ref{fig:cmssm_exclusion} shows the fraction of CMSSM points compatible with the current LHCb measurement and the expected ultimate precision in the 
$(m_{1/2},m_0)$ plane. They are compared to the region excluded at 95\% C.L. by the ATLAS SUSY searches in channels with missing transverse energy (MET) 
obtained on 5.8~fb$^{-1}$ of data at 8~TeV~\cite{ATLAS-CONF-2012-109} and the expected reach of 300~fb$^{-1}$ at 14~TeV~\cite{Abdullin:1998pm}, which shows 
that the sensitivity through the $B_s \to \mu^+ \mu^-$ decay improves approximately as the reach of direct searches. However, while searches in the jets + MET 
channels are directly sensitive to the $m_{1/2}$ and $m_0$ parameters, the $B_s \to \mu^+ \mu^-$ decay probes a complementary region of the CMSSM parameter 
space, accessible to direct searches only through the $H/A \rightarrow \tau \tau$ channel. 

\begin{figure}[t!]
\begin{center}
\includegraphics[width=8.cm]{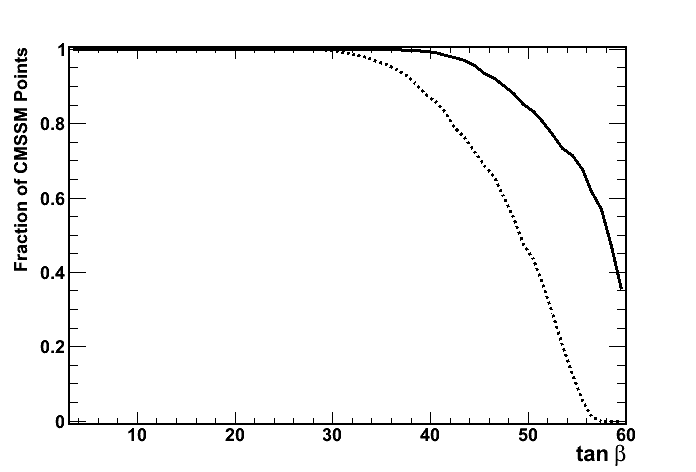}%
\caption{Fraction of CMSSM points obtained through a 4-parameter flat scan passing the LHC SUSY constraints and in agreement with the present BR($B_s \to \mu^+ \mu^-$) 
measurement of Eq.~(\ref{bsmumu_current}) (continuous line), and with the prospective range of Eq.~(\ref{bsmumu_ultimate}) (dotted line), as a function of $\tan\beta$.}
\label{fig:cmssm_fraction}
\end{center}
\end{figure}

We quantify the fraction of the CMSSM points in agreement with the BR($B_s \to \mu^+ \mu^-$) constraint in Fig.~\ref{fig:cmssm_fraction}. As expected the BR($B_s \to \mu^+ \mu^-$) provides us with a powerful constraint for CMSSM points having large values of $\tan\beta$.
The fractions of our generated CMSSM points, for which we also enforce the requirements to have masses of the sfermions below 3.5 TeV, of the gauginos below 3 TeV and of the CP-odd Higgs boson below 2~TeV to make the results directly comparable to those for the pMSSM in the next section, which are compatible with the $B_s \to \mu^+ \mu^-$ rate constraints, are summarised in Table~\ref{tab:cmssm_exclusion}. About 11\% of the CMSSM points not excluded by LHC SUSY searches in our scan are excluded by the current LHCb bound. This fraction increases to 31\% for the estimated final accuracy on the branching ratio of (\ref{bsmumu_ultimate}).
We observe that by restricting the analysis to CMSSM points with large values of $\tan\beta$, {\it i.e.}\ $\tan \beta > 40$, these fractions increase to 21\% and 55\%, 
respectively.  Instead, imposing the anticipated sensitivity of the direct SUSY searches with 300~fb$^{-1}$ at 14~TeV, the fraction of our scan points not excluded by 
the direct searches and incompatible with the projected bounds on $B_s\to \mu^+\mu^-$  decreases from 31\% to 23\%.

\begin{table}[!t]
\begin{center}
 \begin{tabular}{|c|c|c|}
\hline
Fraction of points & Current bounds & Projected bounds\\
\hline
All CMSSM points & 82.7\% & 62.8\% \\
\hline
Accepted CMSSM points & 81.2\% & 61.4\% \\
\hline
Points not excluded by LHC searches & 89.2\% & 69.0\% \\
\hline
 \end{tabular}
\end{center}
\caption{Fraction of CMSSM points compatible with the BR($B_s \to \mu^+ \mu^-$) constraint.\label{tab:cmssm_exclusion}}
\end{table}

\subsection{pMSSM model}
\label{sec:3}

The pMSSM relaxes the correlations introduced by the mass universality assumptions of the CMSSM and allows us to study the inter-relations between the $B_s \to \mu^+ \mu^-$ yields and the MSSM parameters in a general model. Since only a few of these parameters enter in the calculation of the $B_s \to \mu^+ \mu^-$ branching fraction, the pMSSM offers also a viable framework to study 
the complementarity of the constraints from this process with those derived from direct searches by ATLAS and CMS.

\begin{figure}[t!]
\begin{center}
\includegraphics[width=8.cm]{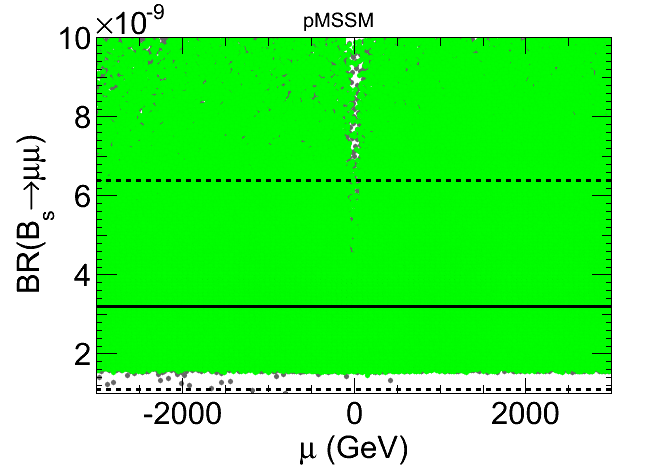}%
\includegraphics[width=8.cm]{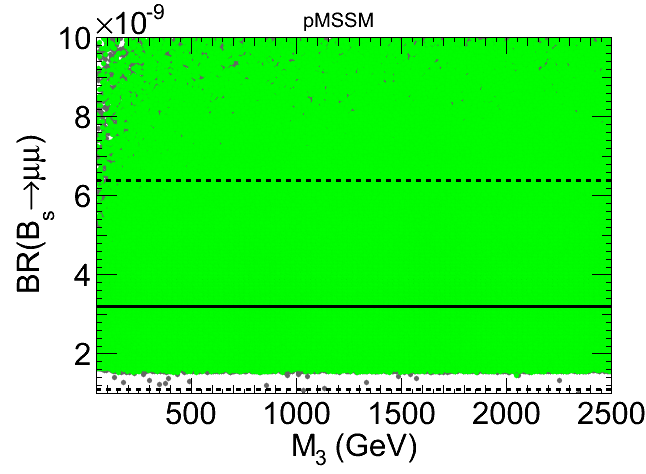}\\[0.1cm]
\includegraphics[width=8.cm]{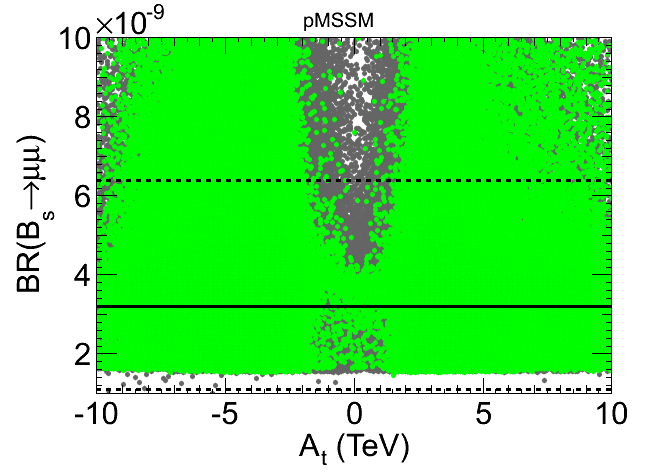}%
\includegraphics[width=8.cm]{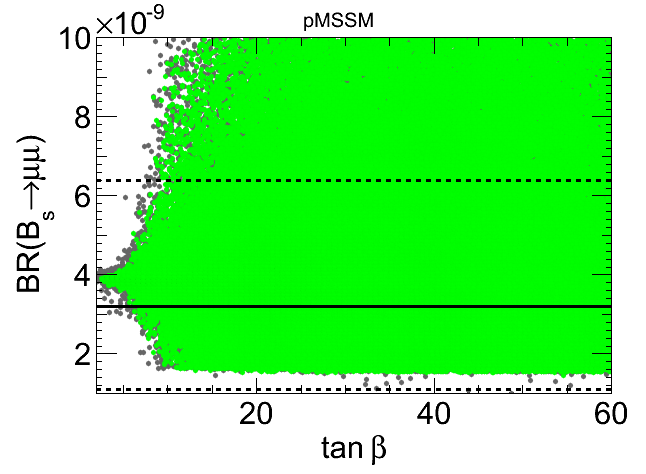}\\[0.1cm]
\includegraphics[width=8.cm]{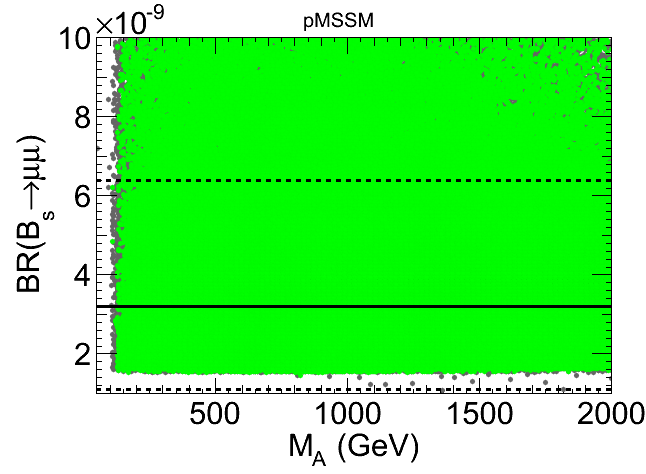}%
\includegraphics[width=8.cm]{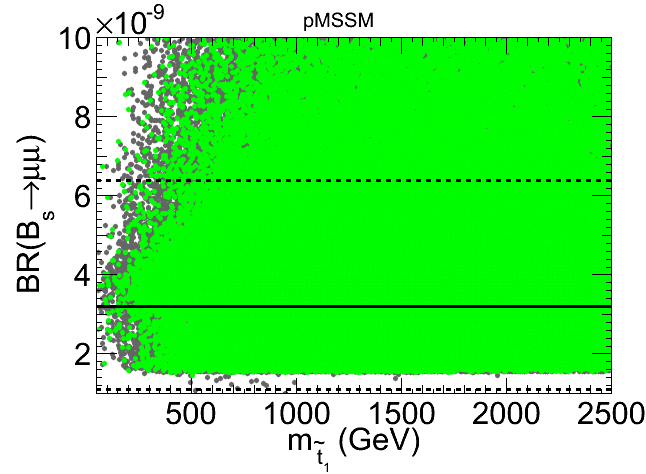}%
\caption{Untagged BR($B_s \to \mu^+ \mu^-$) vs.\ the parameters $\mu$ (upper left), $M_3$ (upper right), $A_t$ (middle left), $\tan\beta$ (middle right), $M_A$ (lower left) and $m_{\tilde t_1}$ (lower right). The solid line corresponds to the central value of the BR($B_s \to \mu^+ \mu^-$) measurement, and the dashed lines to the 2$\sigma$ experimental deviations. The green points are those in agreement with the Higgs mass constraint.}
\label{fig:pmssm}
\end{center}
\end{figure}

The analysis performed here adopts the method and tools described in \cite{Arbey:2011un,Arbey:2011aa}. We perform flat scans of the 19  pMSSM 
parameters in the ranges:
\begin{equation}
\begin{gathered} 
 M_1,M_2 \in [-2500,2500] {\;\rm GeV};\; M_3 \in [50,2500] {\;\rm GeV};\; \tan\beta \in [1,60]\\
M_A \in [50,2000] {\;\rm GeV};\; A_t,A_b,A_\tau \in [-10,10] {\;\rm TeV};\; \mu \in [-3,3] {\;\rm TeV}\\
m_{\tilde{\ell}_{L,R}} \in [50,2500] {\;\rm GeV};\; m_{\tilde{q}_{L,R}} \in [50,3500] {\;\rm GeV}\;.
\end{gathered}
\end{equation}

\begin{figure}[h!]
\begin{center}
\includegraphics[width=9.cm]{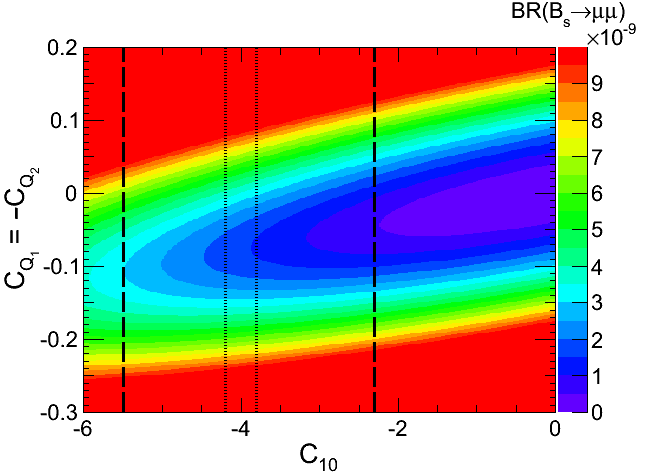}%
\caption{Variation of the untagged BR($B_s \to \mu^+ \mu^-$) in the plane ($C_{10},C_{Q_1}$). The dotted vertical lines delimit the range of 
$C_{10}$ in the CMSSM, and dashed lines the range in the pMSSM.}
\label{fig:pmssm2}
\end{center}
\end{figure}

The dependence of the BR($B_s \to \mu^+ \mu^-$) values calculated at each pMSSM point with the most relevant pMSSM parameters is given in 
Fig.~\ref{fig:pmssm} for all the valid points and those having $123 < M_h < 129$ GeV. Contrary to the case of the CMSSM, here even after 
imposing the Higgs mass constraints a sizeable number of points with a value of BR($B_s \to \mu^+ \mu^-$) below the SM prediction (down to $0.5 \times 10^{-9}$) is obtained. These low values are reached for $\tan\beta \gtrsim 10$ and $m_{\tilde{t}_1} \gtrsim 300$ GeV.
This observation is important for the prospect of improving the lower experimental bound on the decay rate.\footnote{Note that the lower reachable value we obtain is smaller than the one obtained in the recent study of Ref.~\cite{Altmannshofer:2012ks}. This is because we use the full MSSM expressions with no assumption on $C_{10}$, and use non universal masses for the SUSY particles.}

\begin{figure}[h!]
\begin{center}
\includegraphics[width=8.cm]{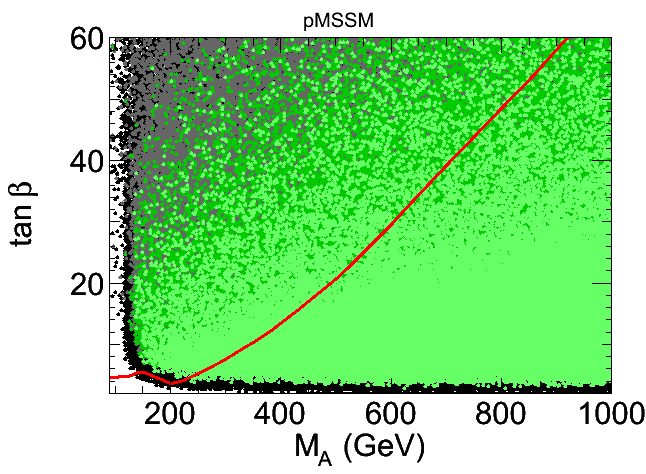}
\includegraphics[width=8.cm]{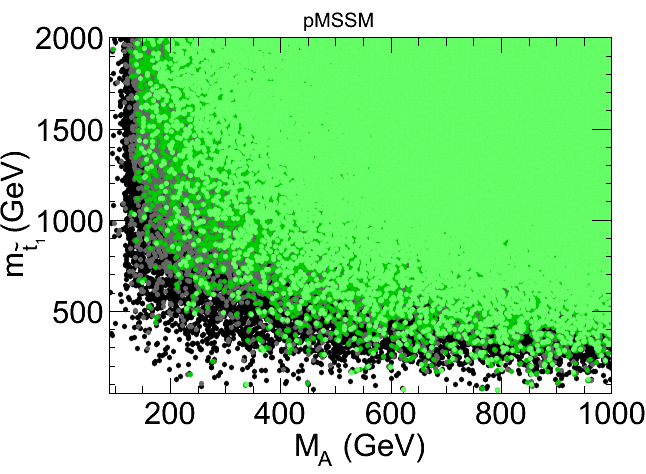}
\caption{Constraints from BR($B_s \to \mu^+ \mu^-$) in the ($M_A,\tan\beta$) and ($M_A,m_{\tilde{t}_1}$) parameter planes. The black points corresponds to all the valid pMSSM points and those in grey to the points for which $123 < M_h < 129$ GeV. The dark green points in addition are in agreement with the latest BR($B_s \to \mu^+ \mu^-$) range given in Eq.~(\ref{bsmumu_current}), while the light green points are in agreement with the prospective LHCb BR($B_s \to \mu^+ \mu^-$) range given in Eq.~(\ref{bsmumu_ultimate}). The red line indicates the region excluded at 95\% C.L. by the CMS $A/H\to\tau^+\tau^-$ searches (from \cite{CMS-HIG-2012-050}).}
\label{fig:pmssm3}
\end{center}
\end{figure}

The BR($B_s \to \mu^+ \mu^-$) dependence on the $C_{10}$ and $C_{Q_{1}}=-C_{Q_{2}}$ Wilson coefficients in the minimal flavour violation (MFV) framework~\cite{D'Ambrosio:2002ex,Hurth:2012jn} is shown in Fig.~\ref{fig:pmssm2}. It is instructive to observe that the values of BR($B_s \to \mu^+ \mu^-$) can decrease down to 0 for $C_{10}=C_{Q_{1}}=0$. However, in the pMSSM, the variation of $C_{10}$ is limited to the interval [-5.0,-2.6], even when applying constraints from $B \to K^*\mu^+\mu^-$ observables, so that the lowest value which can be reached is around $0.5\times10^{-9}$.\footnote{In general non-SUSY MFV scenarios, $C_{10}$ can admit larger ranges, leading to constraints also coming from the lower bound of BR($B_s \to \mu^+ \mu^-$) as shown in~\cite{Hurth:2012vp}.}

The impact of the present and future determinations of BR($B_s \to \mu^+ \mu^-$) on the parameters most sensitive to its rate: ($M_A,\tan\beta$) and ($M_A,m_{\tilde{t}_1}$) is shown in Fig.~\ref{fig:pmssm3}, where we give all the valid pMSSM points from our scan, those with $123 < M_h < 129$ GeV and, highlighted in green, those in agreement with the present BR($B_s \to \mu^+ \mu^-$) range~(\ref{bsmumu_current}) and the ultimate constraint (\ref{bsmumu_ultimate}) at 95\% C.L.
As already discussed in Ref.~\cite{Arbey:2011aa}, the constraints from BR($B_s \to \mu^+ \mu^-$) affect the same pMSSM region, at large values of $\tan\beta$ and small values of $M_A$, also probed by the dark matter direct detection constraints and, more importantly, the $H/A \to \tau^+ \tau^-$ direct Higgs searches at the LHC~\cite{CMS-HIG-2012-050,Aad:2011rv}. The search for the $H/A \to \tau^+ \tau^-$ decay has already excluded a significant portion of the parameter space where large effects on BR($B_s \to \mu^+ \mu^-$) are expected. We also note that the stop sector is further constrained by direct searches in $b$-jets + MET channels, which disfavour small values of $m_{\tilde{t}_1}$. The figure shows that it is difficult for $M_A$ and $m_{\tilde{t}_1}$ to be simultaneously light.

In more quantitative terms, we compute the fractions of all the accepted pMSSM points, of those not excluded by the jets + MET and $H/A \to \tau^+ \tau^-$ searches by ATLAS~\cite{Aad:2011rv} and CMS~\cite{CMS-HIG-2012-050}, and those also compatible at 90\% C.L. with the ATLAS and CMS Higgs data using the analysis of Ref.\cite{Arbey:2012bp}, which are compatible with the (\ref{bsmumu_current}) and (\ref{bsmumu_ultimate}) constraints at 95\% C.L. on $B_s \to \mu^+ \mu^-$. 
\begin{figure}[t!]
\begin{center}
\includegraphics[width=8.cm]{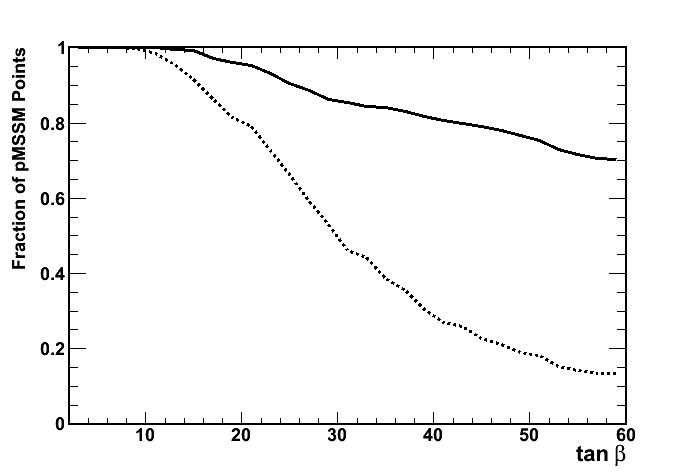}%
\caption{Fraction of pMSSM points passing the LHC SUSY and Higgs mass constraints and in agreement with the latest BR($B_s \to \mu^+ \mu^-$) measurement of Eq.~(\ref{bsmumu_current}) (continuous line), and with the prospective range of Eq.~(\ref{bsmumu_ultimate}) (dotted line), as a function of $\tan\beta$.\vspace*{-0.4cm}}
\label{fig:pmssm_fraction}
\end{center}
\end{figure}
Results are summarised in Fig.~\ref{fig:pmssm_fraction} and Table~\ref{tab:pmssm_exclusion}. The current LHCb result rules out just below 3\% of the pMSSM points compatible with the LHC direct SUSY searches and the Higgs results. The projected bound, assuming the central value coincides with the SM expectation, will increase the reach by an order of magnitude to 30\% of the points and severely constrain solutions with very large values of $\tan \beta$. By then, the direct searches for 
SUSY in channels with jets + MET and $H/A \rightarrow \tau^+\tau^-$ will also have extended their sensitivity to a much larger part of the pMSSM parameter space. Extrapolating the current bounds to 300~fb$^{-1}$, the fraction of our scan points not excluded by the direct searches but excluded by the projected bounds on $B_s\to \mu^+\mu^-$  will decrease from 30\% to $\simeq$20\%. 

If SUSY is indeed realised in nature and a signal from the direct searches at ATLAS and CMS is observed by then, it would be interesting to perform a quantitative test of 
consistency between the mass of the SUSY states being observed and the branching fraction of this, until not long ago, elusive decay. In particular, if the pseudoscalar 
Higgs and the scalar top masses are determined by ATLAS and CMS, the precise value of 
BR($B_s \to \mu^+ \mu^-$) can be used to severely constrain the combination of ($\mu A_t$, $\tan\beta$) in the MSSM.
\begin{table}[t!]
\begin{center}
 \begin{tabular}{|c|c|c|}
\hline
Fraction of points & Current bounds & Projected bounds\\
\hline
All pMSSM points & 95.3\% & 67.8\% \\
\hline
Accepted pMSSM points & 97.7\% & 78.1\% \\
\hline
Points not excluded by LHC searches & 95.1\% & 63.3\% \\
\hline
Points compatible at 90\% C.L. with Higgs results & 97.2\% & 70.0\%\\
\hline
 \end{tabular}
\end{center}
\caption{Fraction of pMSSM points, obtained through a 19-parameter flat scan, compatible with the BR($B_s \to \mu^+ \mu^-$) constraint.\label{tab:pmssm_exclusion}}
\end{table}%
Moreover, by constructing alternative observables such as double ratios of leptonic decays formed from the decays $B_s\to \mu^+\mu^-$, $B_u\to \tau\nu$, $D\to \mu\nu$ and $D_s\to \mu\nu/\tau\nu$, it could be possible to enhance the sensitivity of the individual decays, through the cancellation of hadronic uncertainties, and stronger constraints can be obtained~\cite{Akeroyd:2010qy,Akeroyd:2011kd}.

\section{Conclusions}

The observation of the rare decay $B_s \to \mu^+ \mu^-$ and the first determination of its branching fraction by the LHCb experiment represent a major milestone of the probe of physics beyond the SM through rare decays of $b$ hadrons. 
The excellent agreement of the measured value with the SM prediction has raised the question of its implications on the viability of SUSY. In this paper, we have reviewed the predictions for the branching fraction of this decay in the SM and the MSSM and discussed the impact of the new LHCb result and the expected final LHC accuracy on the SUSY parameter space in two models: the CMSSM and the pMSSM.
We observe that, despite the significant differences between the two models, the sensitivity of the $B_s \to \mu^+ \mu^-$ rate is significant in specific regions of the parameter space of these models, mostly at large values of $\tan \beta$, regions which are also probed by direct SUSY particle searches at ATLAS and CMS. As a result, the constraint derived from the current LHCb result removes $\sim$10\% of the scan points in the CMSSM and a few \% in the pMSSM, which are not already excluded when the bounds from direct SUSY searches and the Higgs data are applied. This is a consequence of the suppression of the SUSY contributions for intermediate $\tan \beta$ values and/or large masses of the 
pseudo-scalar Higgs boson $A$, where the branching fraction in the MSSM does not deviate from its SM prediction. 
The situation in other constrained MSSM scenarios, such as AMSB and GMSB, is similar to that in the 
CMSSM, with high sensitivity at large $\tan \beta$ \cite{Akeroyd:2011kd}.
The improved accuracy on the branching fraction measurement expected from the 14~TeV runs together with the expected improvements in the theory uncertainties, will boost the sensitivity, in particular for the region $\tan \beta > 50$ which could be almost completely constrained, and underline the complementarity of direct and indirect searches for supersymmetry through the possibility of consistency checks, if the heavy Higgs bosons can be observed in the direct searches conducted by ATLAS and CMS.


\section*{Acknowledgements}
We are grateful to Monica Pepe Altarelli for reviewing the text, to Guido Martinelli and Vittorio Lubicz for discussion on 
the expected accuracy of lattice calculations.


\bibliographystyle{h-physrev5}
\bibliography{bsmumu}

\end{document}